\documentclass[a4paper,fleqn,usenatbib]{mnras}

\usepackage[T1]{fontenc}
\usepackage{ae,aecompl}

\usepackage{graphicx}	% Including figure files
\usepackage{amsmath}	% Advanced maths commands
\usepackage{amssymb}	% Extra maths symbols

\usepackage{xhfill}% http://ctan.org/pkg/xhfill
\usepackage{multirow}

\newcommand\eg{{\it e.g.} }

\newcommand\ie{{\it i.e.} }

\newcommand{\ditto}[1][.4pt]{~\textquotedbl~}
\title[ ]{Infrared observations of white dwarfs and the implications for the accretion of dusty planetary material 
 }

\author[A. Bonsor et al.]{
Amy Bonsor$^{1}$\thanks{E-mail: abonsor@ast.cam.ac.uk}, Jay Farihi$^{2}$, Mark C. Wyatt$^{1}$ and Rik van Lieshout$^{1}$ 
\\
$^{1}$Institute of Astronomy, University of Cambridge, Madingley Road, Cambridge, CB3 0HA, UK\\
$^{2}$University College London, Department of Physics and Astronomy, Gower Street, London WC1E 6BT, UK 
}

\date{Accepted XXX. Received YYY; in original form ZZZ}

\pubyear{2015}

\begin{document}
\label{firstpage}
\pagerange{\pageref{firstpage}--\pageref{lastpage}}
\maketitle

\begin{abstract}

Infrared excesses around metal polluted white dwarfs have been associated with the accretion of dusty, planetary material. This work analyses the available infrared data for an unbiased sample of white dwarfs and demonstrates that no more than 3.3\% can have a wide, flat, opaque dust disc, extending to the Roche radius, with a temperature at the disc inner edge of $T_{\rm in}=1,400$K, the standard model for the observed excesses. This is in stark contrast to the incidence of pollution of about 30\%. We present four potential reasons for the absence of an infrared excess in polluted white dwarfs, depending on their stellar properties and inferred accretion rates: i) their dust discs are opaque, but narrow, thus evading detection if more than 85\% of polluted white dwarfs have dust discs narrower than $\delta r <0.04r$, ii) their dust discs have been fully consumed, which only works for the oldest white dwarfs with sinking timescales longer than hundreds of years, iii) their dust is optically thin, which can supply low accretion rates of $<10^7$gs$^{-1}$ if dominated by PR-drag, and higher accretion rates, if inwards transport of material is enhanced, for example due to the presence of gas, iv) their accretion is supplied by a pure gas disc, which could result from the sublimation of optically thin dust for $T_*>20,000$K. Future observations sensitive to faint infrared excesses or the presence of gas, can test the scenarios presented here, thereby better constraining the nature of the material fuelling accretion in polluted white dwarfs.

\end{abstract}


\section{Introduction}

The first white dwarf found to have infrared emission, over and above that predicted for the stellar photosphere, was G29-38 \citep{ZuckermanBecklin1987}. The initial debate in the literature \citep{Wickramsinghe1988,Tokunaga1988, Haas1990, Graham1990, Graham1990a} considered the possibility that the emission could be from a brown dwarf companion. A consensus, however, was quickly reached that the emission resulted from dusty material, based on the gross under-prediction of the observed fluxes by any brown dwarf model \citep{Telesco1990} and the lack of any companion detected in Keck imaging \citep{Kuchner1998}. The current interpretation associates the excess emission with dust accreting onto the white dwarf, linked to pollution and the presence of metal lines in the stellar spectrum \citep{Koester1997}. 

Pollution from elements heavier than helium, whose presence can only be explained by the accretion of external material, are observed for at least 30\% of white dwarfs \citep{Zuckerman03, ZK10, Koester2014}. There is good evidence to suggest that the observed metals originate in an outer planetary system orbiting the white dwarf \citep{Farihi10ism, DebesSigurdsson}. Asteroids (or comets) scattered onto star-grazing orbits are thought to be tidally disrupted and accreted onto the white dwarf \citep{jurasmallasteroid, DebesSigurdsson}. The presence of dusty material, alongside gas in a handful of systems, within the Roche limit around polluted white dwarfs provides evidence of the accretion in progress \citep{Gaensicke06,Gaensicke2007, Gaensicke2008, Melis10}. Transits obscuring the polluted white dwarf WD 1145+017, provide further key evidence that the pollution originates from the accretion of disrupted planetesimals \citep{Vanderburg2015, Croll2015, Rappaport2016, Gaensicke2016}.

Further searches for excess emission in the infrared have found that such dusty emission is always associated with pollution. However, only a few percent of all white dwarfs have excess emission in the infrared \citep{Girven2012, Farihi09, Barber2012, Debes2011}. If the dusty material associated with the infrared excesses fuels the accretion, it is puzzling that some highly polluted white dwarfs have no observed infrared excess (\eg WD 1337+705). Nevertheless, infrared excesses are most common amongst the most highly polluted white dwarfs for example, \cite{Farihi09} find that 50\% of white dwarfs with accretion rates higher than $\dot M > 3\times 10^8$gs$^{-1}$ have an infrared excess.

The standard model most commonly used in the literature \citep{Farihi_review} suggests that the infrared emission results from dusty material in a flat, opaque disc, similar to Saturn's rings \citep{JuraWD03}. This model provides a good fit to the observations \citep[\eg][]{Xu2012, JuraWD03}, however, \cite{Farihi09} note that the observations are, in general, consistent with a single temperature black-body, and some authors have modelled the emission as optically thin rings or halos \citep{Reach05, Reach2009}. Silicate emission features, that must result from optically thin emitting regions, were detected for all six white dwarfs searched for such features \citep{Jura2009}. This led \cite{JuraGD362} to invoke a three-part, warped disc and \cite{Reach2009} a flared disc.

If the observed dusty material fuels the accretion, it is important to consider how it is transported from the observed location (at tens to hundreds of stellar radii) onto the star. Radiative forces, namely Poynting-Robertson drag (PR-drag) will cause dust grains to spiral inwards. \cite{Rafikov1} show that PR-drag in an opaque dust disc can explain the observed accretion rates for all but the most highly polluted white dwarfs. \cite{rafikov2} and \cite{Metzger2012}, therefore, suggest a runaway accretion mechanism due to a coupling of the dust and gas, in order to explain the systems with the highest inferred levels of accretion.

The aim of this work is to use the infrared observations of white dwarfs to explore how dusty material exterior to the white dwarf fuels metal pollution in the atmosphere of the white dwarf. In order to do this, we investigate the near-infrared emission of an unbiased sample of white dwarfs collated from the literature, using {\em Spitzer} and {\em WISE} observations.  We start in \S\ref{sec:sample} by discussing the sample and the cumulative distribution of infrared excesses. This is followed by a discussion, in \S\ref{sec:thick}, of whether the observations are consistent with the standard flat, opaque dust discs. In \S\ref{sec:noexcess} we discuss what supplies the accretion in those polluted white dwarfs without an infrared excess, which leads to our predictions for future observations in \S\ref{sec:predictions} and conclusions in \S\ref{sec:conclusions}.

\begin{figure}
\includegraphics[width=0.48\textwidth]{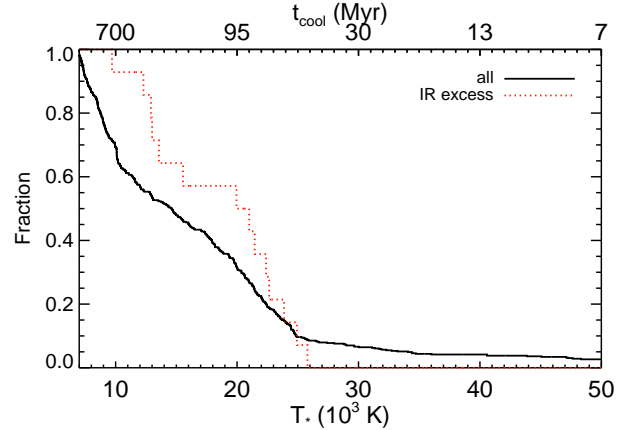}
\caption{The cumulative distribution of stellar effective temperatures and cooling ages, for the full sample, and those stars with an infrared excess. The cooling age is calculated assuming a stellar radius of $0.01R_\odot$. }
\label{fig:tdistribution}
\end{figure}

\begin{table*}
\caption{ Details of the sample. }
\label{tab:survey}
\begin{tabular}{c c c c l l  }\hline\hline
Sample & No. of   & Observatory &No. of & Detections &  Exclusions  \\

& stars & &Detections  & &  \\ \hline\hline

%%%%%%%%%%%%%%%%%%%%%%%%%%%%%%%%%%%%%%%%%%%%%%%%%%%%%%%5
 \cite{Debes_wise} & 276 & \em{WISE} &  4(7)$^{\ddagger}$&  030253.09$-$010833.7$^1$  &    085742.05$+$363526.6$^8$  \\
&&&& 084539.17$+$225728.0$^2$ &  155359.87$+$082131.3$^6$  \\
&&&& 122859.93$+$104032.9$^3$&    024602.66$+$002539.2$^6$ \\
&&&& 161717.04$+$162022.3$^4$&082624.40$+$062827.6$^6$\\

&& &&114758.61$+$283156.2$^5$ &  090611.00$+$414114.3$^6$ \\
 &&&&123432.63$+$560643.0$^5$   & 144823.67$+$444344.3$^6$ \\ 
 &&&&153725.71$+$515126.9$^5$ & 155955.27$+$263519.2$^6$  \\ 

&&&&&  081308.51$+$480642.3$^7$  \\

&&&&& 145806.53$+$293727.0$^8$ \\
&&&&& 011055.06$+$143922.2$^9$  \\
&&&&& 222030.69$-$004107.3$^{10}$ \\\hline

 \cite{Mullally2007}$^*$ & 124 & \em{Spitzer}&  2 &  WD 2326$+$049 & WD 0031$-$274$^6$  \\ 
 &&&&WD 2115$-$560 &  WD 0843$+$358$^{11}$   \\
 &&&&&  WD 1036$+$433$^{11}$ \\

&&&&&  WD 1234$+$481$^{11}$  \\

&&&&&    WD 1616$-$390$^{11}$  \\

&&&&& WD1243+015$^{12}$    \\\hline
%%%%%%%%%%%%%%%%%%%%%%%%%%%%%%%%%%%%%%%%%%%%%%%%%%%%%%%5
\cite{Rocchetto2015} & 134 & \em{Spitzer} &  5 &  WD 1018$+$410 &  \\%$^\dagger$  \\
&&&& WD 2328$+$107\\ &&&&WD 1457$-$086\\ &&&& WD 1015$+$161 \\ &&&& WD 0843$+$516 \\\hline \hline

\end{tabular}

\begin{flushleft}

$^\ddagger$ 4 confirmed with {\em Spitzer}, 7 with {\em WISE} excess in  \citet{Debes_wise}

$^*$ Note $M_*=0.6M_\odot$ is assumed for the \citet{Mullally2007} sample.

{\em Spitzer} observations of \cite{Debes2011} `disc' candidates: $^1$ \cite{Jura2007_dust} $^2$ \cite{Melis10} $^3$  \cite{Brinkworth09, Melis10} 
$^4$ \cite{Brinkworth2012}
$^5$ Identified as a disc in \cite{Debes_wise}
$^6$ UKIDSS or SDSS observations shows clear blending \citep{Barber2014}
$^7$ Background galaxy \citep{wang2014} 
$^8$ Emission in {\em WISE}  is from a background galaxy \citep{farihi08})
$^9$Non-disc like SED Farihi, private communication
$^{10}$Brown dwarf companion \citep{Barber2014} 
$^{11}$ Background object
$^{12}$ Companion
\end{flushleft}

\label{tab:survey}
\end{table*}

\section{The distribution of infrared excesses}
\label{sec:sample}

We aim to assess the presence of dusty material close to white dwarfs (within about a solar radius), where it could fuel the observed accretion. In order to do this we assess the cumulative distribution of infrared excesses, that is the cumulative distribution of white dwarfs with an excess, $\eta_{\lambda}$, above a given level, $f(>\eta_\lambda)$, divided by the number of stars where an excess of the given level could have been detected, where an infrared excess is defined as: 
\begin{equation}
\eta_{\lambda}= \left(\frac{F_{{\rm dust}}}{F_{*}}\right)=\frac{F_{{\rm obs}}-F_{*} }{F_{*}},
\label{eq:excess}
\end{equation}
 where $F_{{\rm obs}}$ is the observed flux at a wavelength $\lambda$, $F_{*}$ is the predicted stellar flux at $\lambda$, and $F_{{\rm dust}}$ is the excess flux, assumed to originate from a dust disc.

In order to assess the cumulative distribution of infrared excesses, we require a large sample of white dwarfs observed in the infrared. We consider {\em Spitzer} and {\em WISE} observations at $4.5\mu$m ({\em W2}), and merge three previous samples, selected without bias towards the level of pollution in the white dwarf atmosphere. This removes the need to de-bias any sample to take into account our ability to detect pollution, which is a strong function of white dwarf temperature. Details of the three surveys are briefly summarised in Table~\ref{tab:survey}.

The full sample contains 528 white dwarfs, of which 14 or 2.6\% have excess emission at 4.5$\mu$m ({\it W2}). The \cite{Debes_wise} sample observed with {\em WISE} suffers from frequent contamination by background sources. We have, therefore, separated their `discs' into two categories; those objects where further observations of the same source with a much smaller field of view (in general with {\em Spitzer}) reveal a similar infrared excess, which are included in the sample and those where further investigations have revealed that the infrared emission originates from another source \eg a background galaxy. Three infrared excesses have neither been refuted, nor observed again, and their status, therefore, remains uncertain. We leave these stars in the full sample for completeness.

 Fig.~\ref{fig:tdistribution} shows the temperature distribution of the merged sample, which is skewed towards hotter white dwarfs. We define $T_*$ as the effective stellar temperature. The majority of stars in our sample have hydrogen-rich atmospheres 459/528, but there is so far no evidence for any differences in the population of pollutants between hydrogen and helium-rich white dwarfs \citep{Wyattstochastic}. This sample has not been methodically searched for metal pollution. A literature search reveals that at least 39 stars in the sample have known pollution (7\%). {\em HST COS} observations of a sub-set (85 stars) find a pollution rate of at least 27\% \citep{Koester2014}, suggesting a similar rate of pollution for the full sample. This compares well to typical pollution rates observed for other samples of white dwarfs.

\begin{figure}
\includegraphics[width=0.48\textwidth]{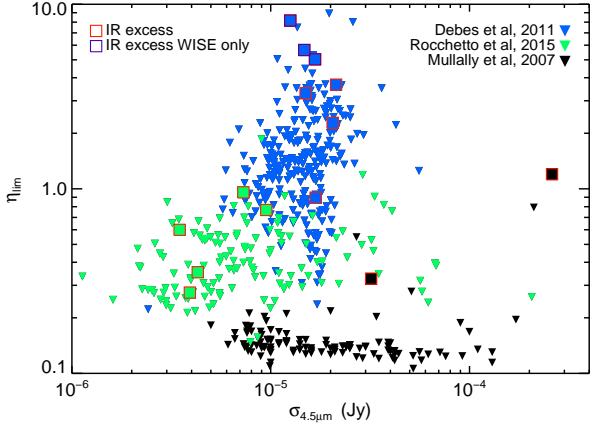}
\caption{The minimum detectable excess, $\eta_{\lim}$ at $4.5\mu$m, as a function of the total error, $\sigma_{4.5\mu {\rm m}}$ for each star in the sample. The red bounded squares show objects where an infrared excess was detected. The three purple bounded squares indicate the infrared excess of the targets for which the infrared excess is based on {\em WISE} observations alone (see \S\ref{sec:sample}). The lack of overlap between the two {\em Spitzer} IRAC samples is partially due to the fact that \citet{Mullally2007} assume a 2\% calibration error for IRAC, while \citet{Rocchetto2015} adopt a more conservative value of 5\%. }
\label{fig:det_thres}
\end{figure}

In order to assess the presence of excess emission in the infrared, the predicted stellar fluxes and associated errors are critical. We follow previous work and use white dwarf atmosphere models \citep{Koester2010}, kindly provided by the author, to fit the observed stellar spectrum in the optical, making use of previously derived stellar effective temperatures. For the \cite{Mullally2007} sample, we use the {\it J}, {\it H} and {\it $K_s$} 2MASS photometry, and effective temperatures presented in their Table 2. For \cite{Debes_wise}, we use the SDSS {\it ugriz} photometry and effective temperatures presented in their Table 1. For \cite{Rocchetto2015}, we use the temperatures quoted in their Table A1 and a mixture of 2MASS, SDSS, APASS and {\em GALEX} photometry, as available for the different stars. The {\em Spitzer} or {\em WISE} observations are then used to calculate the infrared excesses, $\eta_{4.5\mu {\rm m}}$, using Eq.~\ref{eq:excess}. The errors are calculated by summing the quoted observational errors ($\sigma_{\rm obs}$), which are themselves a sum in quadrature of the photometric and calibration errors, and errors on the fit to the stellar spectrum ($ \sigma_{\rm phot}$) in quadrature, such that $\sigma_{4.5\mu {\rm m}}=\sqrt{\sigma_{\rm obs}^2 + \sigma_{\rm phot}^2}$. An excess is considered to be detectable if $F_{{\rm obs}}-F_{*} > 4 \, \sigma_{4.5\mu {\rm m}}$ at $4.5\mu$m. Fig.~\ref{fig:det_thres} shows the minimum detectable excess, given by $\eta_{\rm lim}= 4 \sigma_{4.5\mu {\rm m}}/F_{*}$ as a function of the total error, $\sigma_{4.5\mu {\rm m}}$. Whilst for many stars, excesses as faint as $\eta_{4.5\mu {\rm m}}=0.1$ can be detected, for some of the stars an excess must be larger than $\eta_{4.5\mu {\rm m}}>5$ to be detected. We note here a discrepancy in the assumed calibration of IRAC between \cite{Rocchetto2015}, who assume a conservative value of $5\%$, compared to \cite{Mullally2007} who assume 2\%. We follow the published values, but note that assuming a more realistic 5\% for the \cite{Mullally2007} sample would increase the error bars on our cumulative distribution of infrared excesses, particularly for faint excesses.

\begin{figure}
\includegraphics[width=0.48\textwidth]{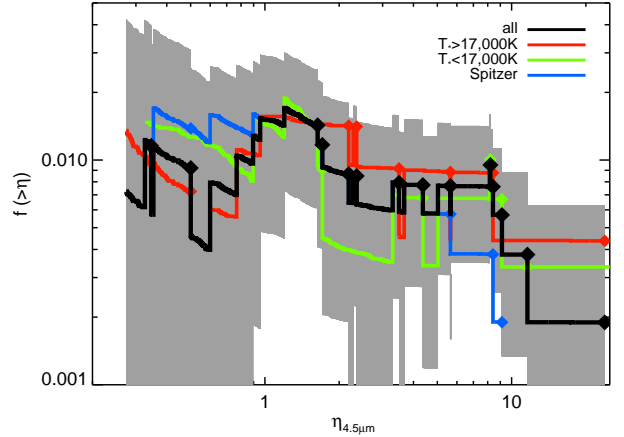}
\caption{The cumulative distribution of infrared excesses at $4.5\mu$m (or {\it W2}), or the number of stars with an excess detection above a given level, divided by the number of stars where such an excess could be detected, as described in \S\ref{sec:sample}, for the full sample (black) and only the {\em Spitzer} confirmed infrared excesses (blue), and those stars in the full sample with $T_*>17,000$K (red) and $T_*<17,000$K (green). The blue line is hidden beneath the black line for faint excesses. Error bars are shown in grey, at $1\sigma$, for the full sample only. }
\label{fig:IF}
\end{figure}

In order to calculate the cumulative distribution of excesses, we consider the number of stars for which a given excess could be detected. The cumulative distribution of infrared excesses is the fraction of those stars for which an excess above the given level is detected. This is the same procedure as used in \cite{Kennedy2013} to assess the distribution of dust in the habitable zone around main-sequence stars and \cite{Wyattstochastic} to assess metal accretion rates onto polluted white dwarfs. The cumulative distribution of excesses for the sample of white dwarfs, at $4.5\mu$m or {\it W2}, is plotted in Fig.~\ref{fig:IF}. The distribution is relatively flat. The grey shaded region shows the $1\sigma$ error bars for the full sample, calculated using binomial statistics, including small number statistics, according to \cite{Gehrels1986}.  The most stringent limit indicated by this plot is the maximum number of stars with an infrared excess above a given level, given by the upper error limit. Plotted in black is the full sample (`all'), which can be compared to the blue dot-dashed line which shows only those objects with additional {\em Spitzer} observations. The red (green) lines show the cumulative distribution of infrared excesses for those stars with $T_*>17,000$K ($T_*<17,000$K). These are both consistent with the full sample, and we find no evidence for significant evolution as the star cools.

\section{flat, opaque dust discs}
\label{sec:thick}

The standard model used in the literature to explain the infrared emission around polluted white dwarfs is a flat, opaque dust disc, situated interior to the Roche limit. We refer the reader to \cite{JuraWD03} for further details of this model. The emission (flux density) from such an opaque dust disc, at a wavelength $\lambda$, is given by: 

\begin{equation}
F_{ {\rm thick}}= \frac{2 \pi \cos (i) }{d^2} \int_{r_{\rm  in}}^{r_{\rm  out}} B_\nu (\lambda, T_{\rm thick})\,  r \, dr, \;
\label{eq:fthick}
\end{equation}
where $B_\nu(T_{\rm thick})$ is the power emitted per unit area per Hertz per solid angle of a black body of temperature $T_{\rm thick}$, $d$ is the distance to the star, $i$ is the inclination of the disc, $r$ is the disc radius, which varies between $r_{\rm  in}$ and $r_{\rm  out}$. The temperature of the disc is assumed to vary as  
 \begin{equation}
T_{{\rm thick}} = \left ( \frac{2}{3 \pi}\right)^{1/4} \left ( \frac{r}{R_{*}}\right)^{-3/4} T_*,
\label{eq:tdisc}
\end{equation}
where $T_*$ is the stellar temperature and $R_*$ is the stellar radius. Our fiducial model considers an opaque, flat dust disc that fills the physical space available to it and extends from $T_{{\rm in}}=1,400$K, which equates to a disc inner radius $r_{\rm  in}$ via Eq.~\ref{eq:tdisc}, to $r_{{\rm out}}=R_\odot$, the Roche radius for a body of density $\rho=3$gcm$^{-3}$, orbiting a star with $M=0.6M_\odot$. If the stellar parameters are not specified, the fiducial model is assumed to orbit a star with $R_*=0.01R_\odot$. We consider variations to the parameters of this fiducial model in \S\ref{sec:disc}.

\subsection{Only nearly edge on flat, opaque dust discs escape detection}
\begin{figure}
\includegraphics[width=0.48\textwidth]{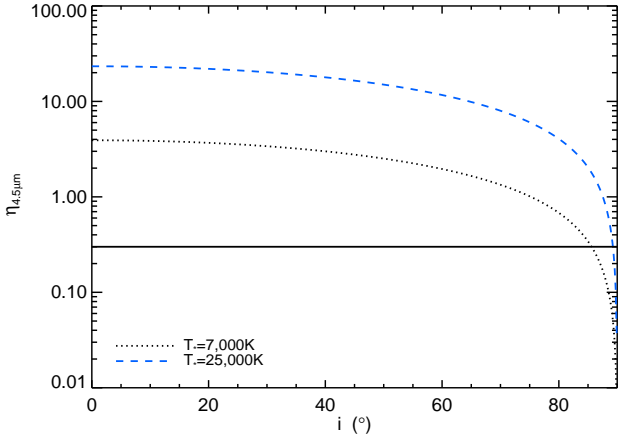}
\caption{The infrared excess of a flat, opaque dust disc as a function of the disc inclination. $\eta_{4.5\mu {\rm m}}$ falls below 30\% for all discs more highly inclined than $i>85^\circ$, for our fiducial disc parameters. }
\label{fig:inc}
\end{figure}

Flat, opaque dust discs are easy to detect in the near-infrared. Our fiducial model would be detected at $\eta_{{\rm lim}}=0.3$, for all discs with $i<85^\circ$, in other words, 90\% of isotropically distributed discs. Fig.~\ref{fig:inc} illustrates the variation in $\eta_{4.5\mu {\rm m}}$ with disc inclination, $i$, for our fiducial model.

\subsection{The cumulative distribution of infrared excesses from opaque dust discs }
\label{sec:infrared_thick}

If opaque, flat dust discs are to explain the observed infrared excesses and supply the accretion in polluted white dwarfs, two conditions must be satisfied. Firstly, the cumulative distribution of infrared excesses must match the observed cumulative distribution of infrared excesses, shown in Fig.~\ref{fig:IF}. We consider this to be satisfied if the fraction of systems with an infrared excess greater than $\eta_{4.5\mu m} = 1$ and $\eta_{4.5\mu m}=8$, match the observed values of $f(>1)=0.015$ and $f(>8)=0.0076$. Secondly, the fraction of systems with a dust disc that leads to detectable pollution must match the observed pollution rate of about $\zeta_{\rm poll}= 30\%$. Opaque dust discs generally accrete at sufficiently high rates to lead to detectable pollution, and therefore, for the current purposes we also equate the fraction of systems with an opaque dust disc to 30\%.

In order to calculate the cumulative distribution of infrared excesses that would result if a fraction $\zeta_{poll}$ of the sample have a flat, opaque dust disc, we place $N=100$ discs, based on our fiducial disc model, around every star in the sample. Each disc is assigned a random inclination based on an isotropic distribution, such that the number of discs with inclination between $i$ and $i + di$ is proportional to $\sin i$. For each star in the sample, we take the effective temperature, $T_*$, stellar radius, $R_*$ and the predicted stellar flux, as calculated in \S\ref{sec:sample}, then the infrared excess at $4.5\mu$m in the model population is calculated as $\eta_{4.5\mu m} = \frac{F_{\rm thick}}{F_*}$, where $F_{\rm thick}$ is the emission from an opaque, flat dust disc calculated using Eq.~\ref{eq:fthick}. The cumulative distribution of infrared excesses is then calculated, taking into account that only a fraction $\zeta_{poll}$ of the sample have a disc. This technique assumes that every star in our sample has an equal probability to have a dust disc, independent of any observed pollution or infrared emission.

\begin{figure}
\includegraphics[width=0.48\textwidth]{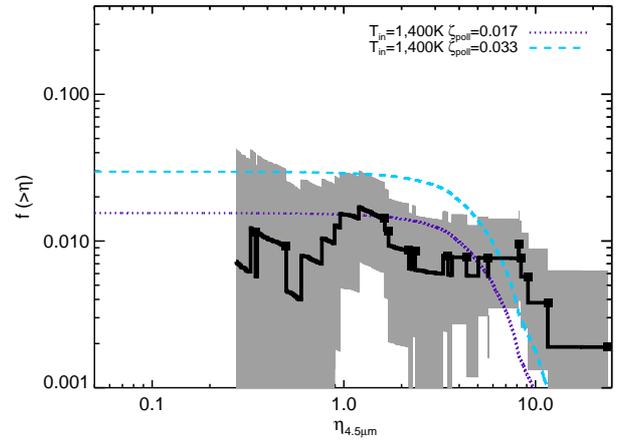}

\caption{The cumulative distribution of infrared excesses at $4.5\mu$m is shown by the thick black line, with black squares for the detected excesses, and $1-\sigma$ errors indicated by the grey shaded region. This is compared to a model, purple dotted line (blue dashed line), in which all stars in the sample have an equal probability to have a flat, opaque dust disc, based on our fiducial model, but only 1.7\% (3.3\%) of the sample are polluted (have a dust disc). }
\label{fig:fdisc_thick}
\end{figure}

The purple dotted line on Fig.~\ref{fig:fdisc_thick} shows the cumulative distribution of infrared excesses from our fiducial model, with $\zeta_{poll}=0.017$, derived such that $f(>1)$ matches the observed value of 0.015. This model provides a reasonable fit to the observed infrared excesses, although a hotter temperature at the disc inner edge would more readily reproduce the largest observed infrared excesses.

However, the low pollution fraction for the cumulative distribution of infrared excesses ($\zeta_{\rm poll} = 1.7\%$) indicates that not all stars in the sample can have an opaque dust disc based on the fiducial model. The blue dashed line shows the cumulative distribution of infrared excesses, with the fraction with $\eta_{4.5\mu {\rm m}}>1$ matched to the $1\sigma$ upper limit on the observed value. This leads us to conclude that up to 3.3\% of the sample could have an opaque, flat dust disc based on the fiducial model. This value remains $<7\%$, taking into account the $3-\sigma$ error limits. In the following section, we investigate whether this conclusion can be altered by changing the parameters of the disc model. The variation in the infrared emission of an opaque, flat dust disc with the stellar parameters of the sample is discussed in the Appendix.

\subsection{The effect of the disc properties}
\label{sec:disc}
\begin{figure}
\includegraphics[width=0.48\textwidth]{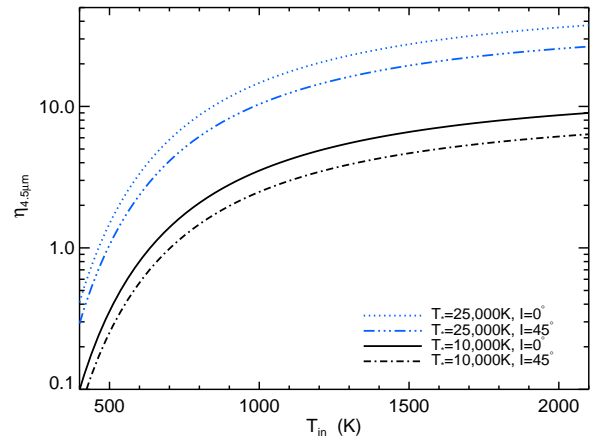}

\caption{The infrared excess, $\eta_{4.5\mu {\rm m}}$, as a function of the temperature at the disc inner edge, for our fiducial disc model, and different stellar temperatures and disc inclinations. }
\label{fig:frac_tin}
\end{figure}

\begin{figure}
\includegraphics[width=0.48\textwidth]{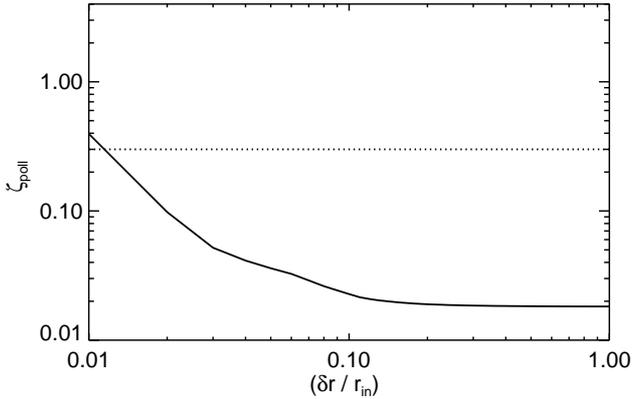}
\caption{The fraction of systems that have detectable pollution, $\zeta_{\rm poll}$, as a function of the disc width, $\delta r$, if the cumulative distribution of infrared excesses is to match the observed distribution at $f(>1)=0.015$. The model discs follow the fiducial disc, except that the disc width, $\delta r$, is varied (see \S \ref{sec:narrow} for further details). The straight dotted line indicates a typical pollution level of 30\%.   }
\label{fig:frac_discwidth}
\end{figure}

\subsubsection{Inner temperature, $T_{\rm in}$}
The emission at $4.5\mu$m is dominated by the hottest dust, thus, the choice of temperature at the disc inner edge can significantly change the infrared emission. Physically, the temperature at the disc inner edge might be determined by sublimation. Fig.~\ref{fig:frac_tin} shows that the emission from opaque dust discs falls off significantly if those discs contain no dust hotter than about 700K. This suggests that the cumulative distribution of infrared excesses could be reproduced by opaque dust discs with a range of inner temperatures, with most stars having a dust disc cooler than 700K. Such cool dust can only lie interior to the Roche limit for stars cooler than $T_* < 12,000$K. Given that the dust is thought to be released during the tidal disruption of planetary bodies interior to the Roche limit, it seems unlikely that most systems would only have dust exterior to the Roche limit. In addition to which, the temperatures for the observed dust discs are generally significantly hotter than 700K \citep[\eg][]{Hoard2013}. We, therefore, conclude that the absence of infrared excesses is unlikely to be explained by the discs being cool.

\begin{figure}
\includegraphics[width=0.48\textwidth]{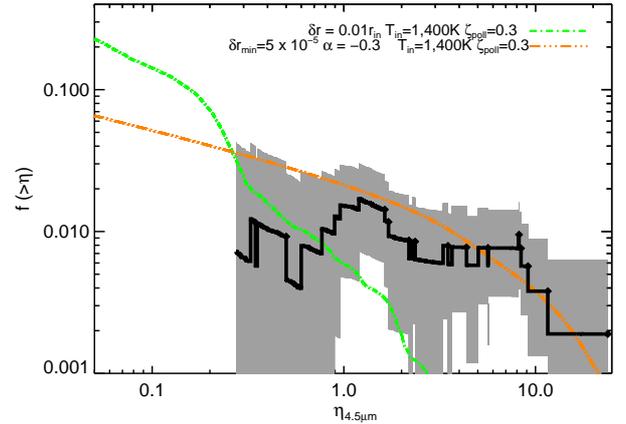}
\caption{The cumulative distribution of infrared excesses at $4.5\mu$m is shown by the thick black line, with black squares for the detected excesses, and $1-\sigma$ errors indicated by the grey shaded region. This is compared to two models in which a fraction, $\zeta_{\rm poll}=30\%$ of stars in the sample have an equal probability to have a narrow flat, opaque dust disc, based on our fiducial model. In the first, green dot-dashed line, all discs have a fixed  width $\delta r=0.013r_{\rm  in}$. In the second, orange dot-double-dashed line, the fraction of discs with a width $\left(\frac{\delta r }{r_{\rm  in}}\right)$ is assumed to vary as $\left(\frac{\delta r }{r_{\rm  in}}\right)^{\alpha}$. Plotted is a best fit value of $\alpha=-0.3$, assuming that disc widths vary from $5\times 10^{-5}r_{\rm  in}$ to $r_{\rm  in}$. } 
\label{fig:fdisc_narrow}
\end{figure}

\subsubsection{Radial width, $\delta r$}
\label{sec:narrow}
Dusty material confined to a narrow belt can escape detection in the infrared \citep{Farihi10}. In order to assess how narrow the dust belts must be in order to match the observed cumulative distribution of infrared excesses, we place a narrow, opaque dust disc, of fixed radial width, $\delta r$, around every star in the sample. The disc inclination is selected from an isotropic distribution and the resulting cumulative distribution of infrared excesses calculated using the same technique as in \S\ref{sec:infrared_thick}. The disc width is defined such that $r_{\rm  out}= r_{\rm  in}( 1+ \left(\frac{\delta r}{r_{\rm  in}}\right))$. The fraction of systems with a dust disc, or detectable pollution, $\zeta_{\rm poll}$, is adjusted such that $f(>1)$ matches the observed value of $f(>1) =0.015$. Fig.~\ref{fig:frac_discwidth} shows the fraction of the sample that must have detectable pollution, $\zeta_{\rm poll}$, as a function of the disc width. If more than $30\%$ of the sample are to be polluted, and all dust discs have the sample radial width, a fixed disc width of less than $\delta r<0.013r_{\rm  in}$ is required to match the infrared observations at $f(<1)$.

Fig.~\ref{fig:fdisc_narrow} shows the observed cumulative distribution of infrared excesses (black solid line) compared to the cumulative distribution of infrared excesses calculated from a model population in which a fraction $\zeta_{\rm poll}$ of the sample have a narrow, flat, opaque dust disc of constant width $\delta r=0.013 r_{\rm  in}$ (green dot-dashed line). This distribution is not consistent with the observed excesses, the largest of which require broader discs. We, therefore, consider a scenario in which the dust discs have a distribution of disc widths, such that the fraction of discs with a fractional width $\frac{\delta r}{r_{\rm  in}}$ is proportional to $\left (\frac{\delta r}{r_{\rm  in}} \right) ^\alpha$. This leads to a model with two parameters, $\alpha$ and the minimum disc width, $\delta r_{min}$. We consider that a physically reasonable minimum disc width is $\delta r_{\rm  min}=10^{-6}$, a disc width of hundreds of meters. The difference between the resulting cumulative distribution of infrared excesses and the observed distribution at $\eta_{4.5 \mu m} = 1$ and $\eta_{4.5 \mu m} =8$, weighted by the errors, was minimized, with $\zeta_{\rm poll} = 0.3$ fixed. The best-fit solution has $\delta r_{\rm  min}=5 \times 10^{-5}$ and $\alpha=-0.3$, and balances achieving a good fit to both $f(>1)$ and $f(>8)$. The orange double-dot dashed line on Fig.~\ref{fig:fdisc_narrow} shows the resulting cumulative distribution of infrared excesses. The maximum disc width that can be detected at $\eta_{4.5\mu m}=1$ is $0.04r_{\rm  in}$. Given that $f(>1)=0.015$, at the very least 28.5\% of the sample must have dust discs narrower than $0.04r_{\rm  in}$, whilst 1.5\% have wider dust discs, if $\zeta_{\rm poll}=30\%$. The best-fit model plotted with $\alpha=-0.3$ and $\delta r_{\rm  min}=5 \times 10^{-5}$ has 80\% of the dust discs have widths less than $0.01r_{\rm  in}$. Such an extreme width distribution seems unlikely unless a mechanism exists that can maintain such narrow discs on sufficiently long timescales to explain the prevalence of pollution.

\begin{figure*}

\includegraphics[width=0.88\textwidth]{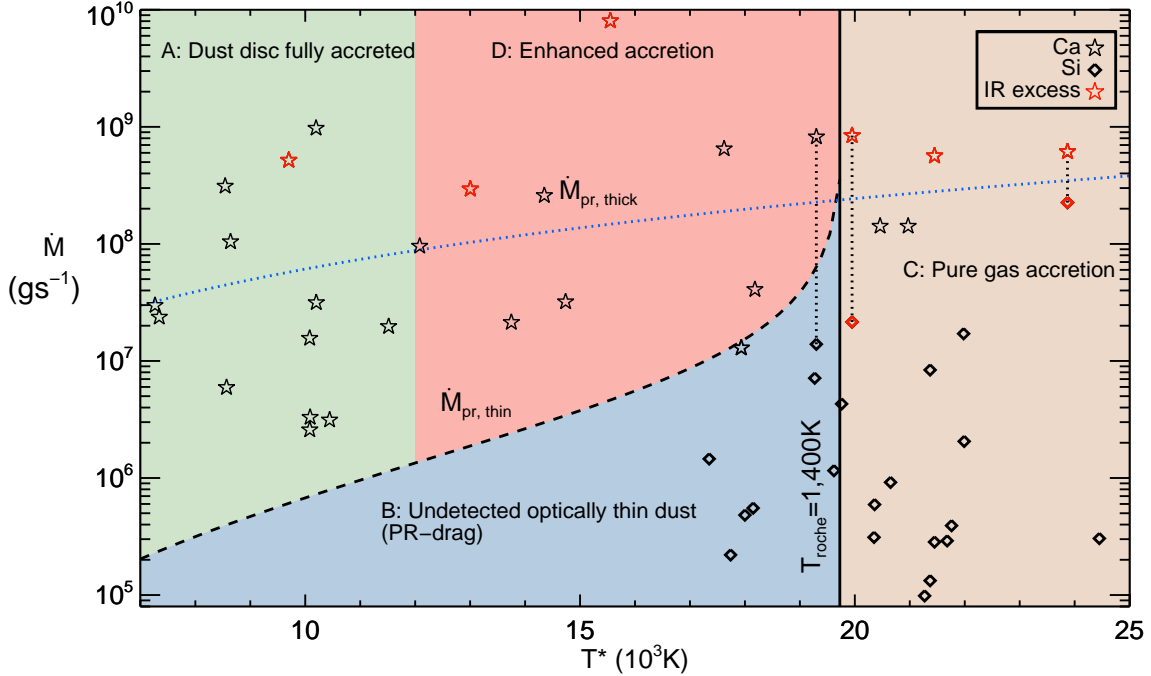}
\caption{The inferred accretion rate of polluted white dwarfs as a function of stellar temperature, indicating the four regions discussed in \S\ref{sec:noexcess} to explain the non-detection of an infrared excess. Region A: Fully accreted dust discs B: Undetected optically thin dust accreted via PR-drag C: Pure gas accretion D: Enhanced accretion of optically thin, undetected dust, or pure gas accretion required to explain the inferred accretion rates. Over plotted are inferred accretion rates from Ca (stars) or Si (diamonds) for all the stars in the sample where pollution has been detected. Red symbols indicate the detection of an infrared excess. The dashed line shows the accretion rate from P-R drag in an undetected optically thin (Eq.~\ref{eq:mdotpr_thin} with $\eta_{4.5\mu m}= 0.3$, $T_{\rm in}=1,400$K and $r_{\rm  out} = r_{\rm  roche}$). For comparison, the blue dotted line shows the PR-drag accretion rate from an optically thick dust disc (Eq.~\ref{eq:mdotpr_thick} dotted) with the parameters of the fiducial model. }

\label{fig:mdot_tstar}
\end{figure*}

\begin{figure}

\includegraphics[width=0.48\textwidth]{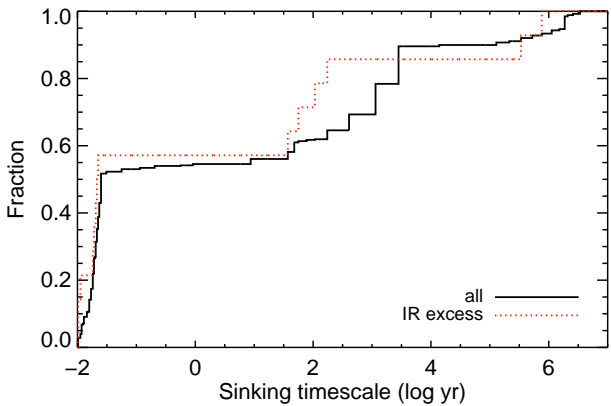}

\caption{The cumulative distribution of gravitational settling timescales for calcium in the white dwarfs in the sample (black), compared to those stars with infrared excesses (red). Settling timescales are calculated using \citet{Koester09}. }
\label{fig:tsink}
\end{figure}

\section{What supplies the accretion in polluted white dwarfs where no infrared excess is detected? }
\label{sec:noexcess}

The aim of this section is to discuss what supplies the accretion for those polluted white dwarfs where no infrared excess is detected. We divide the white dwarfs into four regions, depending on their stellar effective temperature and inferred accretion rate, and discuss what could explain the absence of an infrared excess in each region. Narrow, flat, opaque dust discs can explain the absence of an infrared excess in any region, as long as the distribution of disc widths is strongly peaked towards narrow discs, as discussed in \S\ref{sec:narrow}. Fig.~\ref{fig:mdot_tstar} shows the inferred accretion rates as a function of stellar effective temperature for the sample considered here, with the four regions (A, B, C and D) shaded, green, blue, brown and red. Not all the white dwarfs in the sample have been searched for pollution, and whilst upper limits exist for some objects, only the detections are included on this plot. Plotted for comparison, the dotted blue line on Fig.~\ref{fig:mdot_tstar} also shows the accretion rate from an optically thick, flat dust disc, which \cite{Rafikov1} and \cite{Bochkarev2011} show typically occurs at 
\begin{equation}
\dot{M}_{ \rm PR, thick}=8 \times 10^7 \; \; \left(\frac{R_*}{0.2R_\odot}\right)^2\left(\frac{T_*}{10^4\, {\rm K}}\right)^2\left(\frac{T_{\rm in}}{1400{\rm K}}\right)^2 \; {\rm gs}^{-1}.
\label{eq:mdotpr_thick}
\end{equation}
We also note here that a typical accretion rate for solids from the Inter-Stellar Medium (ISM) would be around $10^4$gs$^{-1}$, \ie below the bottom of the plot. This value uses Eq. 9 of \cite{Farihi10ism}, alongside typical densities and velocities for warm, ionized regions and a factor of 100 to convert from a solar composition. \cite{Farihi10ism} show that the typical accretion rates from the ISM are likely to be a factor of 10-20 times Eddington accretion rates, significantly lower than Bondi-Hoyle accretion rates due to the evaporation of accreting dust grains close to the star.

Observationally, for those white dwarfs where many elements have not been observed, the total accretion rate onto a polluted white dwarf is inferred from observations of a particular species \eg Ca or Si. This requires an assumption regarding the bulk composition of the accreting material. We assume a composition of bulk Earth, as this broadly matches those white dwarfs where the composition has been analysed in further detail \citep{JuraYoung2014}, and provides a consistent base for comparison. The calculated accretion rates are averaged over the sinking timescale in the white dwarf atmosphere of the observed species. We note here the strong detection biases and differences in the observations as a function of white dwarf temperature and atmospheric composition. In particular, a sub-set of the hotter white dwarfs ($T_* > 17,000$K) have been observed by {\em HST}, which has a significantly increased sensitivity to low inferred accretion rates from Si. Such low accretion rates could not have been detected by the ground-based observations that detected Ca, mainly for cooler white dwarfs. Thus, the absence of cool white dwarfs with low inferred accretion rates may be a detection bias, rather than a real feature, as the detection limit for these objects would lie significantly above that for hotter white dwarfs observed with {\em HST}.    
 We also note here the potential biases introduced by the assumed composition, as shown by the large differences in inferred accretion rate from Ca compared to Si, where both observations exist. For the hottest stars radiative levitation of Si can lead to metals in the atmosphere without the need for them to be accreted. \cite{Koester2014} show that this applies to accretion rates of $<10^5$gs$^{-1}$, when converted to a composition of bulk Earth. Accretion rates that could be supported by radiative levitation are not included on Fig.~\ref{fig:mdot_tstar}.

\subsection{A: Dust disc fully accreted (green region)}
\label{sec:lifetime}

Metal sinking timescales are finite, and thus, if and when accretion terminates, metals will persist in the white dwarf atmosphere for up to several sinking timescales. If we observe a star after accretion has terminated, we may detect no infrared excess, despite the fact that the star is polluted. This can provide an explanation for the absence of an infrared excess, particularly for those polluted white dwarfs with long sinking timescales.

 Fig.~\ref{fig:tsink} shows the distribution of gravitational settling timescales for calcium in the sample, taken from Table 5 and 6 of \cite{Koester09}, based on the stellar effective temperature and atmospheric composition. The stellar type classification are taken from \cite{Kleinman2013} for the \cite{Debes2011} sample. 

If we assume that accretion always persists for a fixed time period of $t_{\rm disc}$, that an infrared excess is detectable during this full time period, and that there is an equal probability of observing a given star at any point during its sinking timescale, we can estimate the disc lifetime that would lead to $(f>1)=1.5\%$ of the stars in the sample with an infrared excess, whilst 30\% are polluted. This makes no assumption regarding the nature of the disc. For the full sample, this disc lifetime would have to be $<$15 hours, whereas for those white dwarfs with sinking timescales longer than five hundred years (most of which lie in the green region on Fig.~\ref{fig:mdot_tstar}, although some helium white dwarfs may have higher temperatures), a disc lifetime of $\sim 100$ yrs is consistent with the detection statistics. A disc lifetime of $<$15 hours is significantly shorter than typical estimates for the timescales on which opaque dust discs evolve \citep[\eg][]{Rafikov1, Metzger2012}, and is in contradiction with the multi-epoch observations over decade and longer timescales of some of the earliest known infrared excesses for polluted white dwarfs  (\eg  G29-38 \cite{Graham1990, Reach2009}). The dust, at least in these objects, must last on decadal timescales, or be replenished in approximately steady-state. A disc lifetime of $100$yrs remains short compared to most estimates \citep[\eg][]{Girven2012}, although it is longer, but of the same order of magnitude as the $\sim20$yr disc lifetime estimated by \cite{Wyattstochastic}. Thus, to conclude, whilst a finite dust lifetime cannot explain the absence of an infrared excess for the full population, it provides a good explanation for those white dwarfs with sinking timescales greater than five hundred years.

%;********%%%%%%%%%%%%%%%%%%%%%%%%%%%%%%%%%%%%%%%%%%%%%%%%%%%%%%%%%%%%%%%%%%%%%%%%%%%%%%%%%%%%%%%%%%%%%%%%%%%%%%%%%%%%%%%%%%%%%%%%%%%%%%%%%%%%%%%%%%%%%%%%%%%%%%%%%%%%%%%%%%%%%%%%%%%%%%%%%%%%%%%%%%%%%%%%%%%%%%%%%%%%%%%%%%%%%%%%%%%%%%%%%%%%%%%%

%\begin{figure}

%\includegraphics[width=0.48\textwidth]{mdot_excess.eps}
%\caption{The infrared excess, $\eta_{4.5\mu {\rm m}}$, as a function of accretion rate for those white dwarfs with detections, compared to a theoretical prediction for the infrared excess from optically thin dust, as a function of the accretion rate due to PR-drag, calculated using Eq.~\ref{eq:mdotpr_thin} and Eq.~\ref{eq:fthin}, for a star of temperature $T_*=16,000$K (solid black line) and for every observed accretion rate in the sample (black crosses). The dust is assumed to extend from $T_{\rm in}=1400$K to $r_{\rm out}=0.01R_\odot$. Dust discs with a scale height less than $H=R_*$ become optically thick ($\tau_{||}>0.5$, Eq.~\ref{eq:taup}) for the red dotted portion of the line. The equivalent constant accretion rate from an opaque dust disc with the parameters of the fiducial model ($i<85^\circ$) is plotted as the purple dashed line. }
%\label{fig:mdot_excess}
%\end{figure}

\begin{figure}

\includegraphics[width=0.48\textwidth]{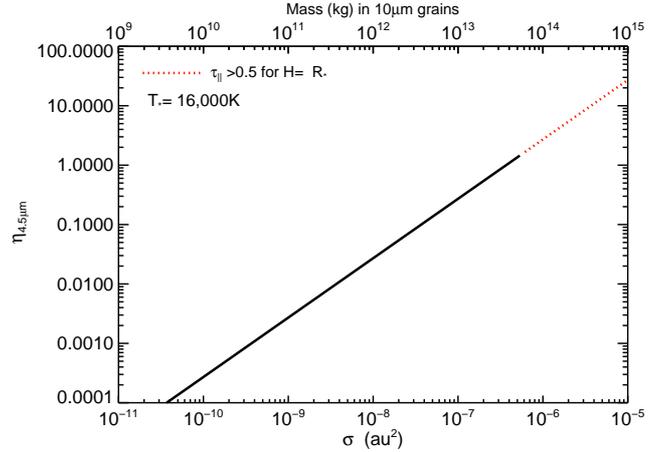}
\caption{The infrared excess, $\eta_{4.5\mu {\rm m}}$, calculated using Eq.~\ref{eq:fthin}, produced by a given cross-sectional area of dust, which relates to the geometrical optical depth by Eq.~\ref{eq:tau}. The dust is assumed to extend from $T_{\rm in }=1,400$K to $r_{\rm out} =0.01R_\odot$, the Roche limit of $0.6M_\odot$ star. The top axis shows the equivalent mass if the cross-sectional area originates entirely from perfectly emitting grains of $10\mu$m diameter and density $\rho=3$gcm$^{-3}$. Dust discs with a height of $H=R_*$ become optically thick 
($\tau_{||}>0.5$, Eq.~\ref{eq:taup}) for the red dotted portion of the line.  }
\label{fig:fdisc_fstar_thin}
\end{figure}

\subsection{B: Undetected optically thin dust accreting by PR-drag (blue region)}
\label{sec:thin}

Optically thin dust can escape detection in the infrared. In order to assess whether optically thin dust can supply the observed accretion, we consider the accretion rate from an optically thin dust disc that is just detectable. The accretion rate of optically thin dust due to PR drag alone is given by  \citep{Whipple1955, Rafikov1, Bochkarev2011, vanLieshout2014a}:
\begin{equation}
\dot{M}_{\rm PR, thin}= \tau \frac{L_*}{c^2},
\label{eq:mdotpr_thin}
\end{equation}
where $\tau$ is the geometrical optical depth of the dust, which is constant in a PR-drag dominated disc. The emission from this dust is given by 
  \begin{equation}
F_{\rm thin}=\frac{2\pi \tau}{d^2}\int^{r_{\rm  out}}_{r_{\rm  in}} B_\nu(\lambda, T_{\rm thin})\,  r\,  dr,    \;
\label{eq:fthin}
\end{equation}
where $T_{\rm thin}$ is the black-body temperature of optically thin dust, given by
\begin{equation}
T_{\rm thin}= \left(\frac{2r}{R_*}\right)^{-1/2} T_*.
\label{eq:tbb}
\end{equation}

The geometrical optical depth of the optically thin dust, $\tau$, is fixed such that the infrared excess $\eta_{4.5\mu m}= \frac{F_{\rm thin}}{F_*}$ lies at a typical detection threshold for this survey of 0.3 (see Fig.~\ref{fig:det_thres}).  The black dashed line on Fig.~\ref{fig:mdot_tstar} shows the accretion rate, via PR-drag, from optically thin dust that is just detectable. Pollution could be supplied by optically thin dust that escapes detection for objects that lie beneath this line. This line moves to higher accretion rates for observations that are less sensitive to faint infrared excesses. For example, undetected optically thin dust could supply the accretion in the three white dwarfs with $T_*$ around 10,000K with accretion rates of about $2 \times 10^6$gs$^{-1}$, which all lie below the equivalent line for their individual detection limits ($\eta_{\rm lim}>2$). The absence of cool white dwarfs with accretion rates lower than this line is in part due to a lack of observations sensitive to such low accretion rates.

Optically thin dust that produces a detectable infrared excess can supply, via PR-drag alone, higher accretion rates than the black dashed line on Fig.~\ref{fig:mdot_tstar}. In fact, optically thin dust can produce an accretion rate higher than a flat, opaque dust disc, via PR-drag, for infrared excesses greater than $\eta_{4.5\mu m}=1$ ($T_*=16,000$K, $T_{\rm in}=1,400$K, $r_{\rm out}=r_{\rm roche}$). However, if the dust resides in a disc with a height, $H$, less than $R_*$, the dusty material becomes optically thick to the incident star-light for infrared excesses greater than 11 (for $T_*=16,000$K), \ie 
$\tau_{||}>0.5$, where
\begin{equation}
\tau_{||}=\tau \frac{(r_{\rm out}-r_{\rm in})}{H},
\label{eq:taup}
\end{equation}
where the disc is assumed to extend, in a similar manner to the fiducial disc model (see \S\ref{sec:thick}) from $T_{\rm in}=1400$K, related to $r_{\rm in}$ by Eq.~\ref{eq:tbb}, to $r_{\rm out}=R_{\odot}$, the Roche limit for a star of mass $0.6M_\odot$.

 The level of infrared emission resulting from optically thin dust should correlate with the accretion rate (Eq.~\ref{eq:mdotpr_thin}, Eq.~\ref{eq:fthin}), whereas on the other hand, if the infrared emission results from opaque dust discs, there should be no correlation between the level of infrared excess and the accretion rate, as the accretion rate depends only on the disc location and stellar properties (see Eq.~\ref{eq:mdotpr_thick}). There is no evidence for such a correlation in the observed infrared excesses, however, this does not prevent optically thin dust from supplying the pollution in those polluted white dwarfs with no detected infrared excess.

%This is compared to the infrared excess as a function of accretion rate from optically thin dust, shown by the black solid line, calculated using Eq.~\ref{eq:mdotpr_thin} and Eq.~\ref{eq:fthin}. Optically thin dust accreted via PR-drag produces larger infrared excesses for higher accretion rates. On the other hand, the blue dashed line shows the infrared excess as a function of accretion rate from an opaque dust disc with an inclination ($i >85^\circ$), calculated using Eq.~\ref{eq:mdotpr_thick} and Eq.~\ref{eq:fthick}. Both calculations assume $T_*=16,000$K and $T_{\rm in}=1400$K. The observed infrared excesses show no evidence for originating from optically thin dust, however, optically thin dust can supply even the highest observed accretion rates, in which case a strong infrared excess results. If the optically thin dust is found in

If optically thin dust is to escape detection, it is important to consider how large the disrupted body could be before it would produce a detectable excess. The calculations so far have focused on the geometrical optical depth of the optically thin dust, $\tau$. We now relate this to the cross-sectional area of dusty material, assuming that the dust extends from $r_{\rm  in}$ to $r_{\rm  out}$, by 
\begin{equation}
\sigma=\int^{r_{\rm  out}}_{r_{\rm  in}} 2 \pi \, r \,\tau \,dr = \pi \tau (r_{\rm  out}^2-r_{\rm  in}^2),
\label{eq:tau}
\end{equation} 
where $r$ is the disc radius. $\sigma$ is a function of the stellar properties, if $r_{\rm in}$ occurs at a fixed temperature, \eg $T_{\rm in} =1400$K, which is related to $r_{\rm in}$ by Eq.~\ref{eq:tbb}. A cross-sectional area, $\sigma$, is equivalent to a mass, $M = \frac{4 }{3 \sqrt{\pi}}\rho \sigma^{3/2}$, if the grains are assumed to be spherical and uniform density. Fig.~\ref{fig:fdisc_fstar_thin} shows the infrared excess as a function of the cross-sectional area of dusty material. This figure shows that a 500m body of density $3$ gcm$^{-3}$ (\ie $10^{10}$kg) disrupted entirely into $10\mu$m grains would not produce a sufficient infrared excess to be detected at $4.5\mu$m ($\eta_{4.5\mu {\rm m}} <0.1$). If the body were disrupted instead into $1$cm grains, a 5km body can escape detection.

If optically thin dust is dominated by PR-drag, it will accrete onto the star on the PR-drag timescale. This timescale can be short, for example, hundreds of years for $100\mu$m grains, or years for $1\mu$m grains around a star with $L_*=0.01L_\odot$ ($T_*=20,000$K). Any optically thin dust must be replenished ( for example from larger grains or fragments of a disrupted asteroid) at a relatively steady rate if there is to be no variability in the observed infrared excess or inferred accretion rates.

\begin{figure}

\includegraphics[width=0.48\textwidth]{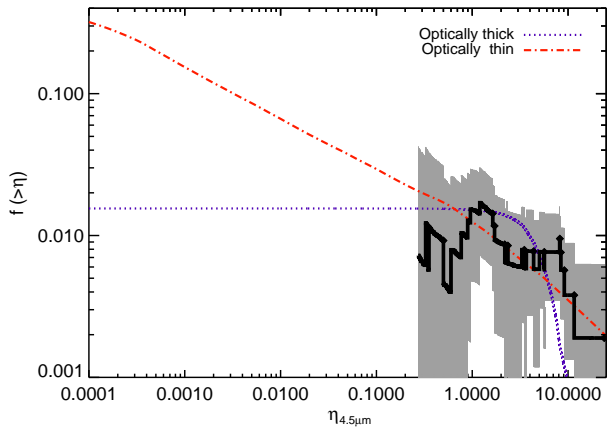}
\caption{The observed cumulative distribution of infrared excess (Fig.~\ref{fig:IF}) compared to a model in which a fraction $\zeta_{\rm poll}=0.017$ of the sample have wide opaque, dust discs (see Fig.~\ref{fig:fdisc_thick}), blue dotted line, and a model in which a fraction $\zeta_{\rm poll}=0.3$ of the sample have optically thin dust with a cross-sectional area taken from the distribution Eq.~\ref{eq:fsigma}, red dot-dashed line. The cumulative distribution of infrared excesses from optically thin dust discs must increase towards faint excesses, if $\zeta_{\rm poll}$ is to equal 30\%.   }
\label{fig:IFthin}
\end{figure}

In order to predict the cumulative distribution of infrared excesses from a population in which all polluted white dwarfs have an optically thin dust disc, we use a cumulative distribution of cross-sectional areas in optically thin dusty material. Given that the purpose of considering optically thin dust is to explain those polluted white dwarfs with no infrared emission, we cannot know the distribution of cross-sectional areas for dusty material, and any choice is somewhat arbitrary. In order to illustrate plausible behaviour, however, we create a distribution of cross-sectional areas that extends to dusty material below the detection limit, based on the observed infrared excesses, noting that there is no evidence to suggest that the observed infrared excesses result from optically thin dust. Every observed infrared excess is converted to an equivalent cross-sectional area of optically thin dust, using Eq.~\ref{eq:fthin} and Eq.~\ref{eq:tau}. The resulting cumulative distribution of cross-sectional areas ($f>\sigma$) is plotted and a best-fit of the form 
\begin{equation}
\log_{10} f(>\sigma) =A + B \log_{10} \sigma.
\label{eq:fsigma}
\end{equation}
determined, where the best-fit parameters are $A=-6.2$ and $B=-0.6$. In order to calculate the cumulative distribution of infrared excesses from optically thin dust, we assumed that every star in the sample had an equal probability to have a cross-sectional area taken from the distribution (Eq.~\ref{eq:fsigma}), extended to $\sigma_{\rm min}=3 \times 10^{-10}$au$^2$. An infrared excess was calculated using Eq.~\ref{eq:excess}, Eq.~\ref{eq:fthin}, Eq.~\ref{eq:tau} and the predicted stellar flux. The same technique as described in \S\ref{sec:sample} and \S\ref{sec:thick} is then used to calculate the cumulative distribution of infrared excesses. Fig.~\ref{fig:IFthin} shows the resulting cumulative distribution of infrared excesses. If the cross-sectional areas in this distribution are converted to accretion rates via PR-drag using Eq.~\ref{eq:mdotpr_thin} and Eq.~\ref{eq:tau}, 30\% of the sample would have, on average, an accretion rate greater than $4\times 10^5$gs$^{-1}$, a low, but typical, detection threshold. In other words this model could plausibly have $\zeta_{\rm poll}=0.3$. The cumulative distribution of infrared excesses from optically thin dust, in this model, increases significantly towards faint excesses. The only way this signature could be imitated by opaque dust discs is if the discs are narrow (see \S\ref{sec:narrow}).

\subsection{C: Pure gas accretion (brown region)} 
We consider the scenario where white dwarf pollution occurs when asteroids scattered interior to the Roche limit are disrupted. If the equilibrium temperature interior to the Roche limit is sufficiently high, dust will sublimate as it is released. This occurs for the hottest white dwarfs ($T_* >20,000$K for $T_{\rm in}=1,400$K), where dust can only survive interior to the Roche limit, for example for those hot white dwarfs with infrared excesses, if it is protected from the stellar radiation by its optical depth or gas partial pressure. Thus, the hottest white dwarfs with an infrared excess are likely to have optically thick dust. We hypothesise that the optical depth of the dust released during the disruption of a planetesimal may depend on the quantity of dust released, as well as the manner in which the disruption proceeds, in particular, how rapidly the dust spreads out, leading to white dwarfs with, and without, dust (infrared excesses). The exact location of this zone depends on the composition of the dust, which determines the temperature at which it sublimates. If 900K is sufficient, this zone could extend to stars as cool as $13,000$K, whereas if only 1,400K is sufficient, this zone starts at 20,000K, as shown on Fig.~\ref{fig:mdot_tstar}.

The accretion of a viscous gas disc proceeds rapidly, depending on the exact parameters of the disc, in particular how actively the disc accretes via the magneto-rotational instability (MRI). The fact that pollution is observed in such a large number of hot white dwarfs with short sinking timescales may require the gas to be replenished, for example, via the disruption of further large fragments.

\subsection{D: Enhanced accretion (red region)} 
\label{sec:enhanced}
It is difficult to explain the absence of an infrared excess for warm white dwarfs (plotted in red for $12,000<T_*<20,000$K), with high accretion rates ($\dot{M}>10^7$gs$^{-1}$). Optically thin dust discs could remain hidden from observations, but would struggle to sustain the observed accretion rates via PR-drag alone. One possibility is that these objects have a different composition such that dust sublimates at lower temperatures, and this region no longer exists (\ie the red region is swamped by the brown region for $T_{in}=900$K). Another possibility is that the high accretion rates in this region are enhanced in an undetected optically thin dust disc by the presence of gas. \cite{rafikov2} and \cite{Metzger2012} suggested a runaway mechanism to enhance accretion rates resulting from drag forces due to gas released by sublimation of dust at the disc inner edge. Such dust, however, would produce a detectable infrared excess, as for optically thin discs, there is insufficient gas released for this mechanism to occur \citep{Metzger2012}. \cite{jurasmallasteroid} and \cite{Xu2012} suggest that gas might be released following high velocity collisions between disrupted bodies or fragments, or sputtering of fragments incident onto a pre-existing dust disc. This gas could enhance accretion rates in an undetected optically thin dust disc, or itself be accreted directly onto the star. Gas has been observed for a handful of white dwarfs which generally have high accretion rates and large infrared excesses \citep{Farihi_review}. However, gas could be present and not yet detected around some (all) white dwarfs, leading to higher accretion rates than from a pure dust disc.

\section{Predictions}
\label{sec:predictions}
Future surveys, that are sensitive to faint infrared excesses, have the potential to detect an increase in the number of white dwarfs that have faint excesses, \ie the cumulative distribution of infrared excesses increases towards faint excesses. Such an increase could be explained by the presence of optically thin dust around a significant proportion of white dwarfs, or alternatively, the presence of narrow ($<0.01r_{\rm  in}$), opaque dust discs. {\em Spitzer IRAC} is already sufficiently sensitive at $4.5\mu$m, however, a larger sample of white dwarfs must be observed in order to reduce the error bars due to small number statistics at faint excesses (see Fig.~\ref{fig:IF}). Future surveys with MIRI on {\em JWST}, using self-calibration, will be able to detect yet fainter excesses.

Variability in either the observed infrared excesses or the metal abundances in polluted white dwarfs would indicate variability in the way accretion proceeds. Variability in the accretion of optically thin dust or pure gas is possible if the dust or gas are not replenished at a steady rate. These variations, however, could also be produced by the accretion of opaque dust discs with a non-flat initial surface density, or coupled to a gas disc \citep{Metzger2012}. Variability in the metal abundances has not currently been definitively detected for any polluted white dwarfs \citep[\eg][]{Debes2008}, although no complete survey exists. A sharp drop in the level of infrared excess has been detected for one object \citep{Xu2014}. 

Future observations that detect the presence of gas for hot white dwarfs with no infrared excess would provide strong evidence in favour of pure gas accretion, or enhanced accretion from gas, for these stars. So far, gas has only been detected for white dwarfs with high accretion rates and strong infrared excesses \citep{Farihi_review}. Hot white dwarfs with high accretion rates provide good targets for future studies searching for emission from gas interior to the Roche limit, but exterior to the radius at which sublimation of dust grains is anticipated.

\section{Conclusions}

\label{sec:conclusions}

In this work, we consider the infrared observations of an unbiased sample of white dwarfs observed with {\em Spitzer} or {\em WISE}. The infrared observations are consistent with the presence of an opaque, flat dust disc, a typical model used to explain the observed excesses, that extends from $T_{\rm in}=1,400$K to the Roche limit, around a maximum of 3.3\% of the sample. This is significantly lower than the pollution rate of around 30\%, and raises the question of what supplies the pollution in those white dwarfs without an infrared excess.

We present {\bf four} potential reasons for the absence of an infrared excess, that depend on the polluted white dwarf's temperature and the total accretion rate, as plotted on Fig.~\ref{fig:mdot_tstar}: 
\begin{enumerate}
\item{The accretion for those stars without an infrared excess could be supplied by narrow, opaque dust discs that escape detection. The cumulative distribution of infrared excesses is only consistent with a distribution of disc widths steeply peaked towards narrow discs. At least 85\% of polluted white dwarfs must have a dust disc narrower than $0.04r_{\rm  in}$.  }
\item {For those white dwarfs with long settling timescales and no infrared excess, an opaque, flat dust disc could have been fully accreted, whilst pollution remains detectable in the white dwarf's atmosphere. For those white dwarfs with sinking timescales longer than five hundred years, this requires a disc lifetime of less than $\sim 100$yrs. }
\item{ For low to moderate accretion rates (up to $< 10^7$ gs$^{-1}$), sparse, optically thin dust, that escapes detection in the near-infrared, could supply the observed accretion via PR-drag alone. In order to sustain the accretion on decadal timescales or longer, this optically thin dust must be replenished. }

 \item{For the hottest white dwarfs ($T_*>17,000$K), dust interior to the Roche limit directly heated by the stellar radiation, sublimates. In order for dust to survive against sublimation, and produce an infrared excess, it must be protected against sublimation by its optical depth. For these hot white dwarfs, the absence of an infrared excess could be explained by pure gas accretion. Pure gas accretion may proceed on short timescales and, therefore, require a continuous supply of material to sustain the high incidence of metals in hot white dwarfs.}

\end{enumerate}
  The absence of an infrared excess for moderately warm polluted white dwarfs, with $12,00K \lesssim T_* \lesssim 17,000K$, with high accretion rates, greater than $> 10^7$gs$^{-1}$, are hard to explain, unless the dust discs are flat, narrow, opaque dust discs, a significant proportion of whom have $\delta r <0.01r_{\rm  in}$, or the accretion occurs from an undetected optically thin dust disc at a rate higher than that from PR-drag alone, potentially linked to the presence of gas.

Future observations that target larger unbiased samples of white dwarfs with {\em Spitzer} IRAC or MIRI on {\em JWST} and are sensitive to faint infrared excesses will constrain whether there is a sharp increase in the cumulative distribution of infrared excesses below $\eta_{4.5\mu {\rm m}} <1$ that would indicate the importance of either optically thin dust, or narrow, opaque dust discs. The detection of gas for white dwarfs without an infrared excess would provide strong constraints on the importance of gas accretion as opposed to dust accretion. Variability in metal abundances would point towards accretion processes with shorter lifetimes, for example driven by optically thin dust or gas accretion.

\section*{Acknowledgements}
AB thanks Grant Kennedy, Marco Rocchetto, James Owen, Dimitri Veras and Siyi Xu for useful discussions that improved the quality of this work. We are grateful to Detlev Koester for comments on the draft manuscript. White dwarf atmosphere models were provided courtesy of Detlev Koester. AB, MCW and RL acknowledge the support of the European Union through ERC grant number 279973. J Farihi acknowledges support from the United Kingdom Science and Technology Facilities Council in the form of an Ernest Rutherford Fellowship (ST/J003344/1). This publication makes use of data products from the Wide-field Infrared Survey Explorer, which is a joint project of the University of California, Los Angeles, and the Jet Propulsion Laboratory/California Institute of Technology, funded by the National Aeronautics and Space Administration.

\section{Appendix: The variation in an infrared excess with stellar properties}
\label{sec:star}
\begin{figure}

\includegraphics[width=0.48\textwidth]{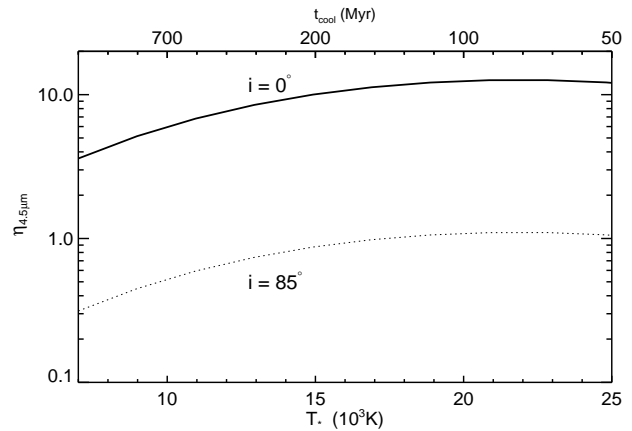}
\caption{The infrared excess from the fiducial disc model as a function of stellar temperature ($T_*$). Cooler stars have fainter infrared excesses, see \S\ref{sec:star} for an explanation. }
\label{fig:ratio_tstar}
\end{figure}

\begin{figure}

\includegraphics[width=0.48\textwidth]{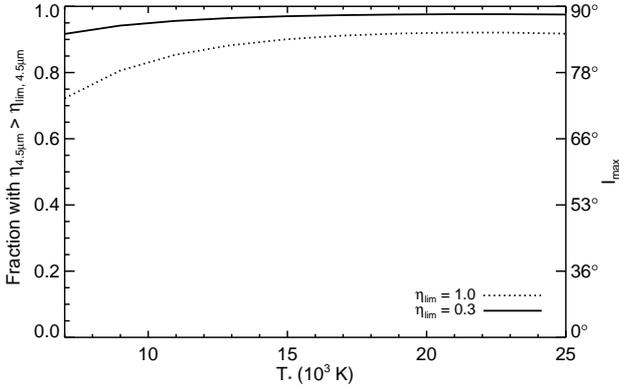}
\caption{The fraction of discs that are detected ($\eta_{4.5\mu {\rm m}} > 0.3$) as a function of the stellar effective temperature. This fraction, $(1-\cos (I_{\rm max}))$, is calculated from the maximum inclination of a disc that can be detected, $I_{\rm max}$, shown on the right-hand y-axis. The discs follow our fiducial parameters.}
\label{fig:frac_tstar_imax}
\end{figure}

 \subsection{Stellar effective temperature, $T_*$}
\label{sec:coolWD}
The level of infrared excess from our fiducial opaque, flat dust disc, falls off as the star cools, as shown on Fig.~\ref{fig:ratio_tstar}. This results from a number of competing effects. The stellar flux, in the Rayleigh-Jeans limit, falls off as $T_*$. The disc flux is dominated by the hottest material, at the disc inner edge. As the sublimation radius decreases, as the star cools, the surface area of the emitting material, and the disc flux, decreases as $r^2$ or $T_*^{8/3}$. This dominates over an increase in the radial width of the disc as the star cools. Thus, the ratio of disc flux to the stellar flux falls off as $T_*^{5/3}$.

The decrease in $\eta_{4.5\mu m}$ with decreasing stellar temperature means that opaque flat discs are harder to detect around cooler stars, as shown by Fig.~\ref{fig:frac_tstar_imax}.  However, even at $T_*=7,000$K, 85\% of discs have $\eta_{4.5\mu {\rm m}}>0.3$, for the fiducial parameters. The temperature distribution of stars in the sample is taken account in calculating the cumulative distribution of infrared excesses plotted in Fig.~\ref{fig:fdisc_thick}, and therefore, cannot explain the absence of infrared excesses compared to pollution.

\bibliographystyle{mn}

\bibliography{ref}

%%%%%%%%%%%%%%%%%%%%%%%%%%%%%%%%%%%%%%%%%%%%%%%%%%

% Don't change these lines
\bsp	% typesetting comment
\label{lastpage}
\end{document}